\documentclass[showpacs,twocolumn,tightenlines,prd]{revtex4-1}
\usepackage{graphicx}
\usepackage{amsmath,amsfonts,amssymb}
\numberwithin{equation}{section}

\graphicspath{{figures/}} 

\usepackage{hyperref}
\hypersetup{
    colorlinks=true,
    linkcolor=blue,
    filecolor=magenta,      
    urlcolor=blue,
    citecolor=magenta,
}

\usepackage{color}

\usepackage{comment}
\def\comment#1{}

\begin{document}

\title{Thermodynamics of Eliashberg theory in the weak-coupling limit}

\author{Sepideh Mirabi, Rufus Boyack, F. Marsiglio}
\affiliation{Department of Physics \& Theoretical Physics Institute, University of Alberta, Edmonton, Alberta T6G~2E1, Canada}

\begin{abstract}
The weak-coupling limits of the gap and critical temperature computed within Eliashberg theory surprisingly deviate from the BCS theory predictions by a factor of $1/\sqrt{e}$. Interestingly, however, the ratio of these two quantities agrees for both theories. 
Motivated by this result, here we investigate the weak-coupling thermodynamics of Eliashberg theory, with a central focus on the free energy, specific heat, and the critical magnetic field. 
In particular, we numerically calculate the difference between the superconducting and normal-state specific heats, and we find that this quantity differs from its BCS counterpart by a factor of $1/\sqrt{e}$, for all temperatures below $T_{c}$. 
We find that the dimensionless ratio of the specific-heat discontinuity to the normal-state specific heat reduces to the BCS prediction given by $\Delta C_{V}(T_{c})/C_{V,n}(T_c)\approx1.43$. This gives further evidence to the expectation that all dimensionless ratios tend to their ``universal values" in the weak-coupling limit. 
\end{abstract}

\pacs{}
\date{\today}
\maketitle

\section{Introduction}
The thermodynamic properties of superconductors~\cite{Kleiner_Buckel} are interesting macroscopic quantities that afford insight into the excitation spectrum, the pairing gap, and also the nature of heat transfer in superconductors. 
The specific heat is one such quantity of great interest due to the fact that, for a second-order, mean-field-like phase transition, it exhibits a discontinuity as the temperature is decreased towards the critical temperature. 
In addition, the presence of a pairing gap leads to an exponential suppression in the low-temperature behaviour of the specific heat, in contrast to the linear temperature dependence of a normal-state metal~\cite{Biondi1958,Geilikman1958}.
Indeed, experimental measurements~\cite{Wang2001} of the specific heat can provide a diagnostic on the importance of strong-coupling corrections and elucidate the nature of the gap. 
In the context of the Eliashberg theory~\cite{Eliashberg1960} of superconductivity, the specific heat has primarily been addressed in the context of the strong-coupling limit~\cite{Marsiglio86,Bennemann}.
Motivated by recent work~\cite{Marsiglio2018,Mirabi2020} elucidating the differences in the gap and critical temperature for weak-coupling Eliashberg theory and BCS theory, here we investigate the thermodynamics of Eliashberg theory in the weak-coupling limit. 

To determine the specific heat, which is proportional to the second derivative of the free energy with respect to temperature, one must first calculate the free energy. In Ref.~\cite{Luttinger1960}, Luttinger and Ward developed the requisite theoretical formalism for calculating the thermodynamic potential of an interacting system. In general, the thermodynamic potential can be calculated by summing all bare closed-loop diagrams~\cite{Luttinger1960}. However, summing such a perturbative expansion proves to be difficult~\cite{AGD}. Luttinger and Ward showed that a partial resummation could be performed by constructing a functional consisting of all closed-loop skeleton diagrams with the bare Green's function replaced by the full irreducible Green's function. 
An expression for the thermodynamic potential can then be calculated, and the first-order variation of this function, with respect to the self energy, vanishes for a Green's function obeying Dyson's equation. The thermodynamic potential is thus self-consistent, and an additional important facet of this approach is that a self energy constructed in this manner ensures the satisfaction of macroscopic conservation laws~\cite{Baym1962}.

The free energy~\cite{Eliashberg1963} for an interacting electron-phonon system~\cite{Eliashberg1960} was then given by Eliashberg. However, the expression is computationally intractable. Nevertheless, with the aid of several plausible assumptions, the momentum integration appearing in this expression can be performed, and by considering the difference between the superconducting and normal-state free energies, a far more tractable expression can be obtained. 
Bardeen and Stephen~\cite{Bardeen1964} derived such a formula, and its utility lies in the fact that it only requires performing a summation over Matsubara frequencies; 
thus, once the mean-field parameters are determined, the thermodynamic properties of superconductors are readily amenable to calculation.
The result in Ref.~\cite{Bardeen1964} is equivalent to a less rapidly-convergent expression obtained by Wada in Ref.~\cite{Wada1964}.

In Ref.~\cite{Bardeen1964} the analysis was confined to the case of an isotropic Fermi surface with a constant density of states. The free-energy difference for a system with an energy-dependent density of states was derived in Ref.~\cite{Mitrovic1983} by Mitrovic and Carbotte; 
these authors also included the effects from the Coulomb interaction and (nonmagnetic) impurities. The extension of the Luttinger-Ward formalism to more general Eliashberg-type theories has also recently been achieved~\cite{Chubukov2005,Benlagra2011}.

The modification of the critical magnetic field due to strong electron-phonon coupling was studied in Ref.~\cite{Rainer1973}. 
Numerical analysis of the free-energy difference for strong-coupling Eliashberg theory was performed by Daams and Carbotte~\cite{Daams1981}. 
In Ref.~\cite{Marsiglio1989}, Marsiglio et. al computed the specific heat in the asymptotic limit of Eliashberg theory, as the interaction strength approaches infinity.
Using the Bardeen-Stephen formula, the condensation energy of a superconductor with a generalized pairing interaction was studied by Tsoncheva and Chubukov~\cite{Tsoncheva2005} from weak to strong coupling strengths. 
Further discussion on the numerical and experimental analysis of superconductors, within the Eliashberg framework, can be found in Ref.~\cite{Bennemann}. For a recent review of when Eliashberg theory is valid, see Ref.~\cite{Chubukov2020}.

In this paper we consider the thermodynamics of weak-coupling Eliashberg theory. The motivation for this limit is due, in part, to the surprising  (and not widely appreciated) result~\cite{Karakozov1976,Dolgov2005,Wang2013,Marsiglio2018,Mirabi2020} that the weak-coupling limit of
Eliashberg theory does not reduce to BCS theory. Indeed, in the weak-coupling limit, the critical temperature $T_{c}$ and the zero-temperature limit of the gap function $\Delta_{0}$ both have corrections of $1/\sqrt{e}$ in comparison to their BCS counterparts. 
Nevertheless, the ratio of these two quantities was known~\cite{Mitrovic1984} early on to limit to the BCS prediction. 
In Ref.~\cite{Marsiglio2018} one of the present authors considered the weak-coupling Eliashberg theory on the imaginary axis, and in Ref.~\cite{Mirabi2020} we extended this analysis to the real axis, thus obtaining a complete understanding of 
the gap and renormalization functions for weak-coupling Eliashberg theory. Here we complement these studies and investigate the specific heat and critical magnetic field. 

The outline of the paper is as follows. In Sec.~\ref{sec:Theory} we use the Bardeen-Stephen formula to study the zero-temperature and critical-temperature limits of weak-coupling  Eliashberg theory.
We obtain the expected corrections to the BCS results; however, the dimensionless ratios agree. 
The numerical calculations of the specific heat and the critical magnetic field are then discussed in Sec.~\ref{sec:Numerics} and finally Sec~\ref{sec:Conclusion} presents the conclusion.

\section{Theoretical Analysis}
\label{sec:Theory}
Determining the free energy in the superconducting phase, within the Eliashberg theory framework, is a non-trivial endeavour.
However, the Bardeen-Stephen formula~\cite{Bardeen1964} provides a useful method to numerically compute the difference between the superconducting and normal-state free energies. 
This formula is given by
\begin{align}
\label{eq:FreeEnergy}
    \frac{\Delta F}{N(0)}\nonumber&=-\pi T\sum_{i\omega_{m}}\left(\sqrt{\omega_{m}^{2}+\Delta^{2}(i\omega_{m})}-\left|\omega_{m}\right|\right)\\
    &\quad\times\left(Z_{s}(i\omega_{m})-Z_{n}(i\omega_{m})\frac{\left|\omega_{m}\right|}{\sqrt{\omega_{m}^{2}+\Delta^{2}(i\omega_{m})}}\right).
\end{align}
Here, $T$ is the temperature, $N(0)$ is the single-spin electronic density of states at the Fermi energy, and $\Delta(i\omega_{m})$ is the frequency-dependent gap function where $\omega_{m}=(2m-1)\pi T$, with $m\in\mathbb{Z}$, is a fermionic Matsubara frequency. 
The renormalization factors in the superconducting and normal states are respectively defined as $Z_{s}(i\omega_{m})$ and $Z_{n}(i\omega_{m})$. 
Natural units $\hbar=k_{B}=1$ are used throughout the manuscript. An important feature of this expression is that it requires knowledge of the gap function only on the imaginary frequency axis. 
Thus, the subtleties~\cite{Mirabi2020,Marsiglio88,Marsiglio2020} involved in analytic continuation to the real-frequency axis are absent. In the case where the gap function is frequency independent, as in BCS theory, Eq.~\eqref{eq:FreeEnergy} reduces to the BCS result~\cite{Bardeen1957} for the free energy difference. One point to keep in mind is that the $\Delta$ appearing here is the self-consistent gap function, that is, in this expression $F$ is not an arbitrary functional of $\Delta$ and as a result $d\Delta F/d\Delta$ is not zero. 

The Eliashberg equations~\cite{Bennemann} consist of the following coupled equations for the superconducting gap $\Delta(i\omega_{m})$ and the normal and superconducting renormalization factors $Z_{n}(i\omega_{m})$ and $Z_{s}(i\omega_{m})$:
\begin{align}
Z_{n}(i\omega_{m})&=1+\frac{\pi T}{\omega_{m}}\left(\lambda+2\sum_{n=1}^{m-1}\lambda(i\nu_{n})\right).\label{eq:Zn}\\
Z_{s}(i\omega_{m})&=Z_{n}(i\omega_{m})+\frac{\pi T}{\omega_{m}} \sum_{m^\prime = -\infty}^\infty\lambda(i\omega_{m}-i\omega_{m'})\nonumber\\
&\quad\times\left(\frac{\omega_{m^{\prime}}}{\sqrt{\omega_{m^{\prime}}^{2}+\Delta^{2}(i\omega_{m^\prime})}}-{\rm sgn}(\omega_{m^{\prime}})\right).\label{eq:Zs}\\
\Delta(i\omega_{m})Z(i\omega_{m})&=\pi T\sum_{m^\prime = -\infty}^\infty\lambda(i\omega_{m}-i\omega_{m'})\frac{\Delta(i\omega_{m^\prime})}{\sqrt{\omega_{m^\prime}^{2}+\Delta^{2}(i\omega_{m^\prime})}}.\label{eq:Delta}
\end{align}
The bosonic Matsubara frequencies are $\nu_{n}=2n\pi T$, where $n\in\mathbb{Z}$, and the electron-phonon coupling is 
\begin{equation}
  \lambda(i\omega_{m}-i\omega_{m^{\prime}})=\frac{2A\omega_{E}}{\omega_{E}^{2}+(\omega_{m}-\omega_{m^{\prime}})^{2}}.
\end{equation}
Here, $\omega_{E}$ is the Einstein frequency and $A=\lambda\omega_{E}/2$ is the weight of the spectral function, where $\lambda>0$ is a fixed interaction strength.
In the next two subsections we obtain an expression for the free-energy difference of weak-coupling Eliashberg theory in the zero-temperature limit and near the critical temperature. 
These expressions will both differ from the BCS results by a factor of $1/e$.

\subsection{Zero-temperature limit}
\label{sec:FreeEnergyZeroTemp}
In the zero-temperature limit, that is, in the limit $\overline{T}=T/\omega_{E}<<1$, the Matsubara frequency summation becomes an integration according to the prescription~\cite{AGD}
\begin{equation}
 \overline{T} \sum_{i\omega_{m}}\rightarrow{\int_{-\infty}^{\infty}}\frac{d\overline{\omega}}{2\pi}.
\end{equation}
The weak-coupling limit is defined~\cite{Marsiglio2018} by $T_{c}/\omega_{E}\ll1$; thus, for any $T<T_{c}$, in this limit one also has $T/\omega_{E}\ll1$, and therefore weak-coupling is synonymous with the zero-temperature limit. In the weak-coupling limit, the zero-temperature gap-function, on the imaginary frequency axis, can be approximated as~\cite{Marsiglio2018}:
\begin{equation}
\label{eq:Gap} \overline{\Delta}\left(i\omega_{m}\right)=\frac{\overline{\Delta}_{0}}{1+\overline{\omega}_{m}^{2}},
\end{equation}
where $\overline{\Delta}_{0}=\Delta_{0}/\omega_{E}$ (in general $\overline{Q}=Q/\omega_{E}$) is the gap parameter, determined by the condition $\Delta_{0}=\text{Re}\Delta(\omega=\Delta_{0})$~\cite{Parks1}.
In Eq.~\eqref{eq:FreeEnergy}, we use the above approximation for $\Delta$, however, we drop the denominator. In the region of small frequencies this is permissible, since the denominator is near unity, and for large frequencies this is also valid since the $O(\omega_{m}^2)$ term under the square root in Eq.~\eqref{eq:FreeEnergy} dominates. In the weak-coupling limit $Z_{s}\approx Z_{n}\approx1+\lambda\equiv Z_{\lambda}$~\cite{Marsiglio2018}. We first define a dimensionless free-energy difference $\Delta f$ by 
\begin{equation}
\Delta f=\frac{\Delta\overline{F}}{N(0)\omega_{E}Z_{\lambda}}.
\end{equation}
Using this definition, Eq.~\eqref{eq:FreeEnergy} then reduces to
\begin{align}
\label{eq:FreeEnergyZeroTemp}
    \Delta f&=-\int_{0}^{\infty}d\overline{\omega}\left(\sqrt{\overline{\omega}^{2}+\overline{\Delta}_{0}^{2}}-\overline{\omega}\right)\left(1-\frac{\overline{\omega}}{\sqrt{\overline{\omega}^{2}+\overline{\Delta}_{0}^{2}}}\right)\nonumber\\
    &=\lim_{L\rightarrow\infty}\left.\left(\overline{\omega}^2-\overline{\omega}\sqrt{\overline{\omega}^{2}+\overline{\Delta}_{0}^{2}}\right)\right|_{0}^{L}\nonumber\\
    &=-\frac{1}{2}\overline{\Delta}_{0}^{2}. 
\end{align}
In Ref.~\cite{Mirabi2020} it was shown that the gap parameter $\overline{\Delta}_0$ is given in the weak-coupling limit as
\begin{equation}
\overline{\Delta}_{0}=\frac{2}{\sqrt{e}}\exp{\left(-\frac{1+\lambda}{\lambda}\right)}=\frac{1}{\sqrt{e}}\Delta_{0,\text{BCS}},\quad \overline{T}\rightarrow0.
\end{equation}
Here, we define $\Delta_{0,\text{BCS}}$ to be the zero-temperature limit of the BCS gap function with renormalization effects included~\cite{Marsiglio2018}.
Thus, the weak-coupling free-energy difference is
\begin{equation}
\label{eq:fZeroTemp}
\Delta f=\frac{1}{e}\Delta f_{\text{BCS}},\quad \overline{T}\rightarrow0.
\end{equation}
The right-hand side of this equation means that Eq.~\eqref{eq:FreeEnergy} is calculated with the BCS gap used for the gap function.
This result can be easily understood as follows. The low-temperature limit of the free-energy difference is the square of the gap function~\cite{AGD} and, as previously shown~\cite{Mirabi2020}, since the weak-coupling limit of the gap function has a factor of $1/\sqrt{e}$ different from the BCS limit, the free-energy difference thus acquires a prefactor $1/e$ in comparison to the BCS limit. In Sec.~\ref{sec:Numerics} we shall numerically confirm this result. 

\subsection{Critical-temperature limit}
In the limit $T\rightarrow{T_{c}}$ the gap function satisfies $\Delta(i\omega_{m})\ll T_{c}$. By performing a small $\Delta/T_{c}$ expansion in Eq.~\eqref{eq:FreeEnergy}, the free-energy difference can be expanded in powers of $\Delta$ as follows:
\begin{equation}
 \frac{\Delta F}{N(0)Z_{\lambda}}\rightarrow-\frac{\pi T_{c}}{4}\sum_{i\omega_{m}}\frac{\Delta(i\omega_{m})^{4}}{\left|\omega_{m}\right|^{3}},\quad T\rightarrow T_{c}.
\end{equation}
As in Sec.~\ref{sec:FreeEnergyZeroTemp}, the gap function can be approximated as $\Delta(i\omega_{m})\approx\Delta_{0}$. In Ref.~\cite{Mirabi2020} it was numerically proved that, for weak-coupling Eliashberg theory, 
$\Delta_{0}(T)/\Delta_{0}(T\rightarrow0)$ is in good agreement with the BCS ratio for this quantity. The temperature dependence of the gap parameter, as $T\rightarrow T_{c}$, can thus be approximated in the weak-coupling limit by the BCS result~\cite{AGD,FetterWalecka}:
\begin{equation}
\frac{\Delta_{0}(T)}{T_{c}}=\pi \sqrt{\frac{8}{7\zeta(3)}}\sqrt{1-\frac{T}{T_{c}}}.
\end{equation}
Note that the weak-coupling factor of $1/\sqrt{e}$ would be present in the numerator and denominator of the left-hand side of this expression, and thus it drops out from this ratio. 
After performing the Matsubara frequency summation~\cite{AGD}, the free-energy difference becomes
\begin{equation}
\Delta f=-\frac{4\pi^2}{7\zeta(3)}\left(1-\frac{T}{T_{c}}\right)^{2}\overline{T}_{c}^2.\label{eq:FreeEnergyTc}
\end{equation}
The analytical approximation for $T_{c}$ in the weak-coupling limit was determined in Ref.~\cite{Marsiglio2018} to be
\begin{equation}
\label{eq:Tc}
T_{c}=\frac{2e^\gamma}{\pi\sqrt{e}}\exp{\left(-\frac{1+\lambda}{\lambda}\right)}=\frac{1}{\sqrt{e}}T_{c,\text{BCS}},
\end{equation}
where $\gamma\approx0.5772$ is the Euler-Mascheroni constant. 
Here, we define $T_{c,\text{BCS}}$ to be the critical temperature for the BCS gap function with renormalization effects included~\cite{Marsiglio2018}.
Thus, the weak-coupling free-energy difference is
\begin{equation}
\Delta f=\frac{1}{e}\Delta f_{\text{BCS}},\quad T\rightarrow T_{c}.
\end{equation}
Combining this equation with the result in Eq.~\eqref{eq:fZeroTemp} we find that, in both the zero-temperature and critical-temperature limits,  the weak-coupling free-energy difference has a correction of $1/e$ compared to the BCS result. 
Since the weak-coupling limit is synonymous with the zero-temperature limit, this explains why the $1/e$ correction factor in the free-energy difference is expected to persist for all temperatures $T<T_{c}$:
\begin{equation}\Delta f=\frac{1}{e}\Delta f_{\text{BCS}},\quad 0\leq \overline{T}\leq \overline{T}_{c}.
\end{equation}
Here we have analytically confirmed this result for $\overline{T}\rightarrow0$ and $\overline{T}\rightarrow \overline{T}_{c}$, and in the next section we numerically confirm this for a range of intermediate temperatures. 
Another way to understand why the correction factor $1/e$ between weak-coupling Eliashberg theory and BCS theory free-energy differences is the same for all temperatures 
is due to the fact that both the zero-temperature gap parameter and $T_{c}$ receive the same weak-coupling correction. That is, the two pertinent energy scales in the respective limits have the same weak-coupling correction, and thus the same correction appears at all temperatures. 

The specific heat (at constant volume) is defined as 
\begin{equation}
C_{V}=-T\left(\frac{\partial^{2}F}{\partial T^{2}}\right)_{V}.\label{eq:SpecificHeat}
\end{equation}
The specific-heat difference is $\Delta C_{V}=C_{V,s}-C_{V,n}$, where the normal-state specific heat is~\cite{Bennemann}: $C_{V,n}(T)=\frac{2\pi^{2}}{3}Z_{n}N(0)T$. 
Using Eqs.~\eqref{eq:FreeEnergyTc} and \eqref{eq:SpecificHeat} we find that $\Delta C_{V}\sim T_{c}$, as $T\rightarrow T_{c}$, and since the weak-coupling result for $T_{c}$ in Eq.~\eqref{eq:Tc} has a prefactor of $1/\sqrt{e}$, the specific-heat difference 
also has a factor of $1/\sqrt{e}$ different from the BCS result. However, the normalized change in the specific-heat difference $\left.\Delta C_{V}/C_{V,n}\right|_{T\rightarrow T_{c}}$ reduces to the BCS result
\begin{equation}
\left.\frac{\Delta C_{V}}{C_{V,n}}\right|_{T\rightarrow T_{c}}=\frac{12}{7\zeta(3)}\approx 1.43.
\end{equation}
Thus, the normalized specific-heat difference in the weak-coupling theory limits to the BCS prediction, despite the fact that $T_{c}$ and $\Delta_{0}$ receive $1/\sqrt{e}$ corrections. As alluded to earlier, we expect that this agreement persists for all temperatures $T<T_{c}$. 
In the numerical analysis presented in the next section this will be verified.

\section{Numerical Analysis}
\label{sec:Numerics}
\subsection{Specific heat}

\begin{figure}[t]
\centering
\includegraphics[width=8.25cm]{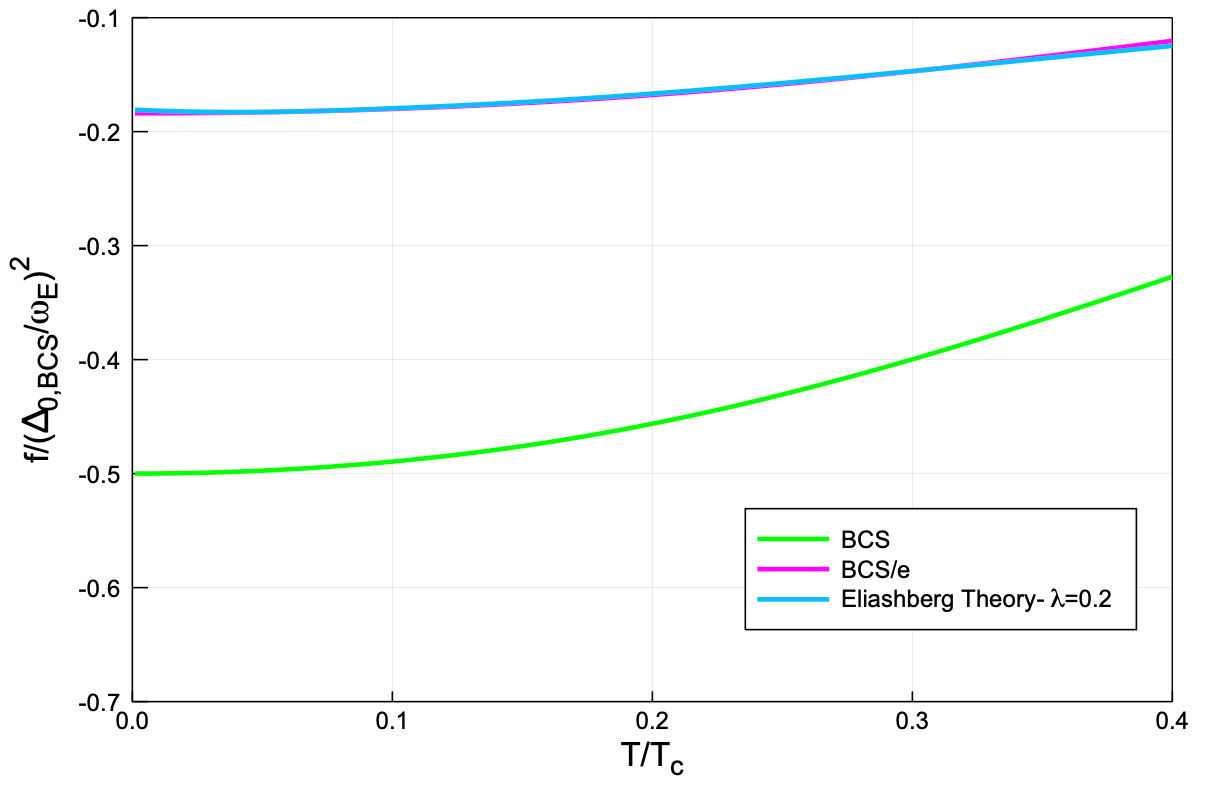}
\caption{A plot of $f/\overline{\Delta}_{0,\text{BCS}}^{2}$ versus $T/T_{c}$. The self-consistent equations~\eqref{eq:Zn}-\eqref{eq:Delta} are solved for $\Delta(i\omega_{m}), Z_{s}$, and $Z_{n}$, which are then inserted into Eq.~\eqref{eq:FreeEnergy}. The blue curve corresponds to Eliashberg theory with $\lambda=0.2$, whereas the green curve is the BCS result, and the fuschia curve is the BCS result multiplied by $1/e$. 
There is very good agreement between Eliashberg theory and $1/e$ times the BCS result.}
\label{fig:Freeenergy}
\end{figure}

\begin{figure}[t]
\centering
\includegraphics[width=8.25cm]{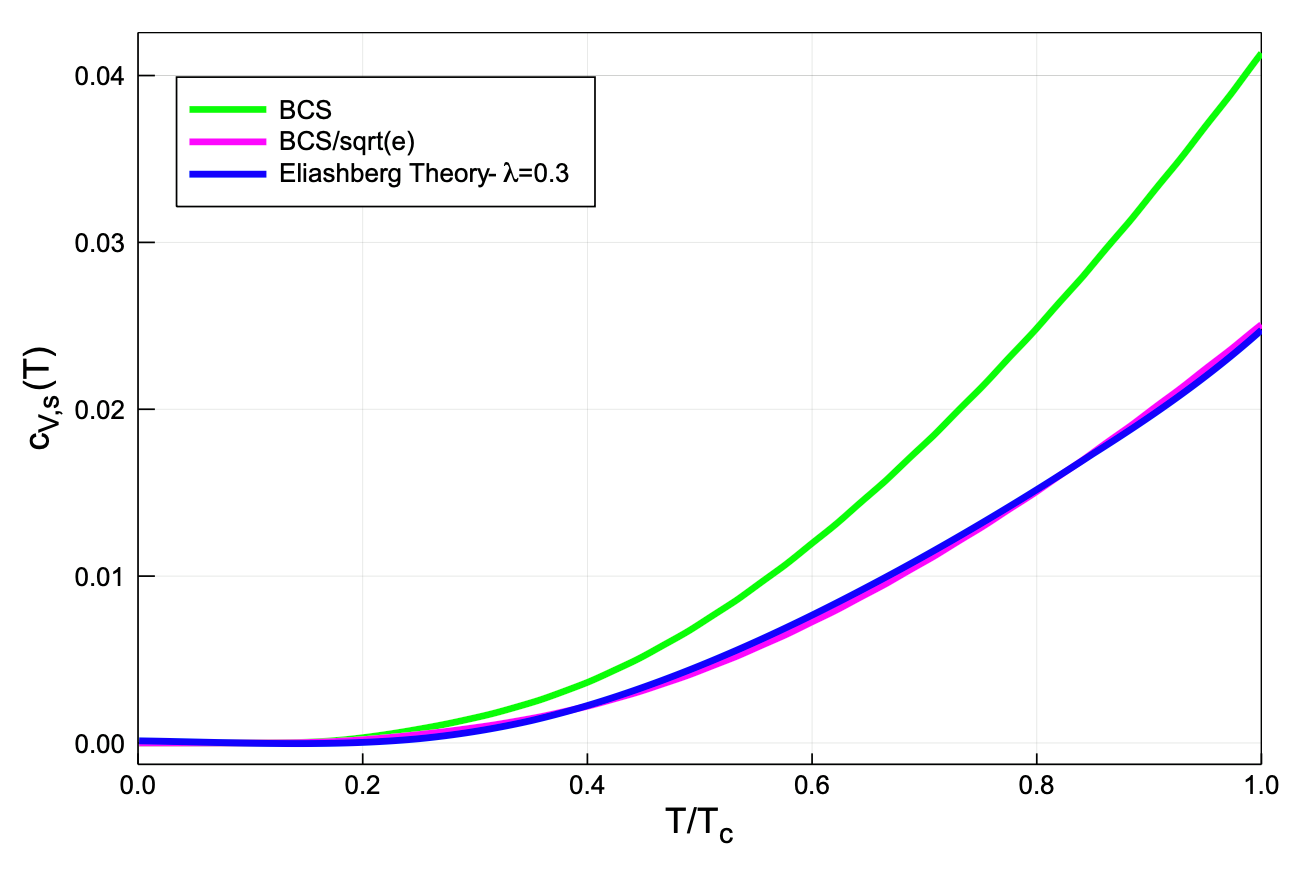}
\caption{A plot of $c_{V,s}(T)\equiv C_{V,s}(T)/(\frac{2\pi^2}{3}Z_{\lambda}N(0))$ versus $T/T_{c}$. For the $\lambda=0.3$ Eliashberg plot (blue), $\overline{T}_{c}=0.009923$. For the BCS plot (green) we set $\overline{T}_{c}=0.009923\sqrt{e}\approx0.03976$, which is valid in the weak-coupling approximation.}
\label{fig:SpecificHeat0}
\end{figure}

In Fig.~\ref{fig:Freeenergy} we plot $\Delta f/\overline{\Delta}_{0,\text{BCS}}^2$ versus $T/T_{c}$ for Eliashberg theory with $\lambda=0.2$, for BCS theory, and also for $1/e$ times the BCS result.
From Eq.~\eqref{eq:FreeEnergyZeroTemp}, the zero-temperature limit of this quantity is expected to be $-1/2$ for BCS theory, whereas for weak-coupling Eliashberg theory the zero-temperature limit is expected to be corrected from the BCS result by $1/e$, namely $-1/(2e)\approx-0.184$. 
As shown in the figure, the numerically computed free-energy difference for weak-coupling Eliashberg theory clearly exhibits a $1/e$ correction compared with the BCS result, for all temperatures $0\leq \overline{T}\leq\overline{T}_{c}$.

\begin{figure}
\centering
\includegraphics[width=8.25cm]{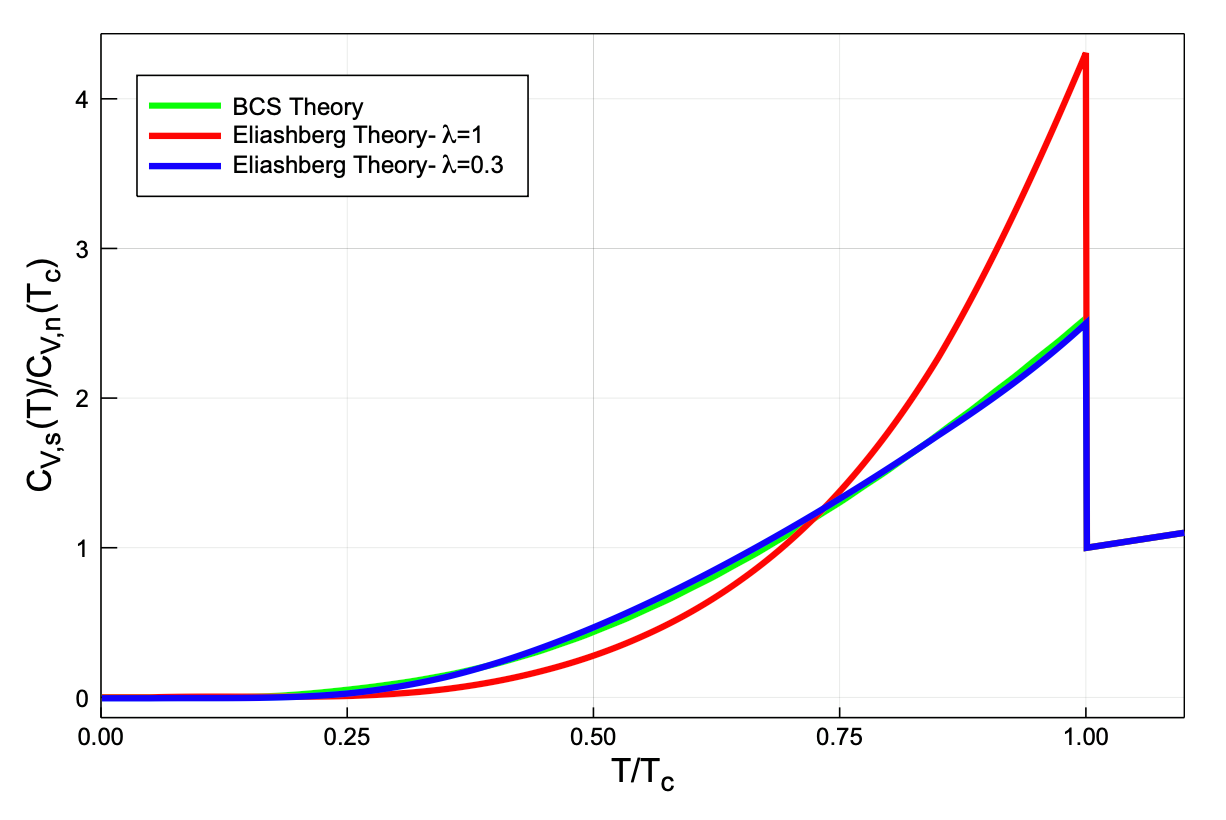}
\caption{A plot of $C_{V,s}(T)/C_{V,n}(T_{c})$ versus $T/T_{c}$. The Eliashberg theory results correspond to coupling constants $\lambda=1$ (red) and $\lambda=0.3$ (blue) whereas the BCS result (with $\Delta$ determined by solving the BCS gap equation) is given in green.}
\label{fig:SpecificHeat1}
\end{figure}

In Fig.~\ref{fig:SpecificHeat0} we plot the superconducting specific heat $C_{V,s}(T)$, normalized by the quantity $\frac{2\pi^2}{3}Z_{\lambda}N(0)$, versus $T/T_{c}$ for the $\lambda=0.3$ Eliashberg theory and for BCS theory. 
For this choice of coupling constant, the Eliashberg case has~\cite{Mirabi2020} $\overline{T}_{c}=0.009923$; for the BCS case we use $\overline{T}_{c}=0.009923\sqrt{e}\approx0.03976$. 
This figure verifies the analytical analysis of the previous section; near $T=T_{c}$ we find that $C_{V,s}\sim\overline{T}_{c}$, and thus the respective specific heats in the Eliashberg and BCS theories differ by a factor of $1/\sqrt{e}$. 
This difference of $1/\sqrt{e}$ appears for all temperatures. To illustrate this fact in another fashion we proceed as follows. 
In Fig.~\ref{fig:SpecificHeat1} we plot the ratio of the superconducting and normal-state specific heats, $C_{V,s}(T)/C_{V,n}(T_{c})$, versus $T/T_{c}$ for the Eliashberg and BCS theories. At the critical temperature there exists a discontinuity in $C_{V}$, which illustrates the occurrence of a second-order phase transition, and $C_{V,s}$ is noticeably larger than $C_{V,n}$. The specific heat ratio for Eliashberg theory with $\lambda=0.3$ and BCS theory shows excellent agreement. However, for Eliashberg theory with $\lambda=1$, we observe a deviation from the BCS result. One should bear in mind that while this specific heat ratio agrees for the two theories, the specific heat itself $(C_{V,s})$ exhibits a $1/\sqrt{e}$ difference between weak-coupling Eliashberg theory and BCS theory, as evinced in Fig.~\ref{fig:SpecificHeat0}.

\begin{figure}[h]
\includegraphics[width=8.25cm]{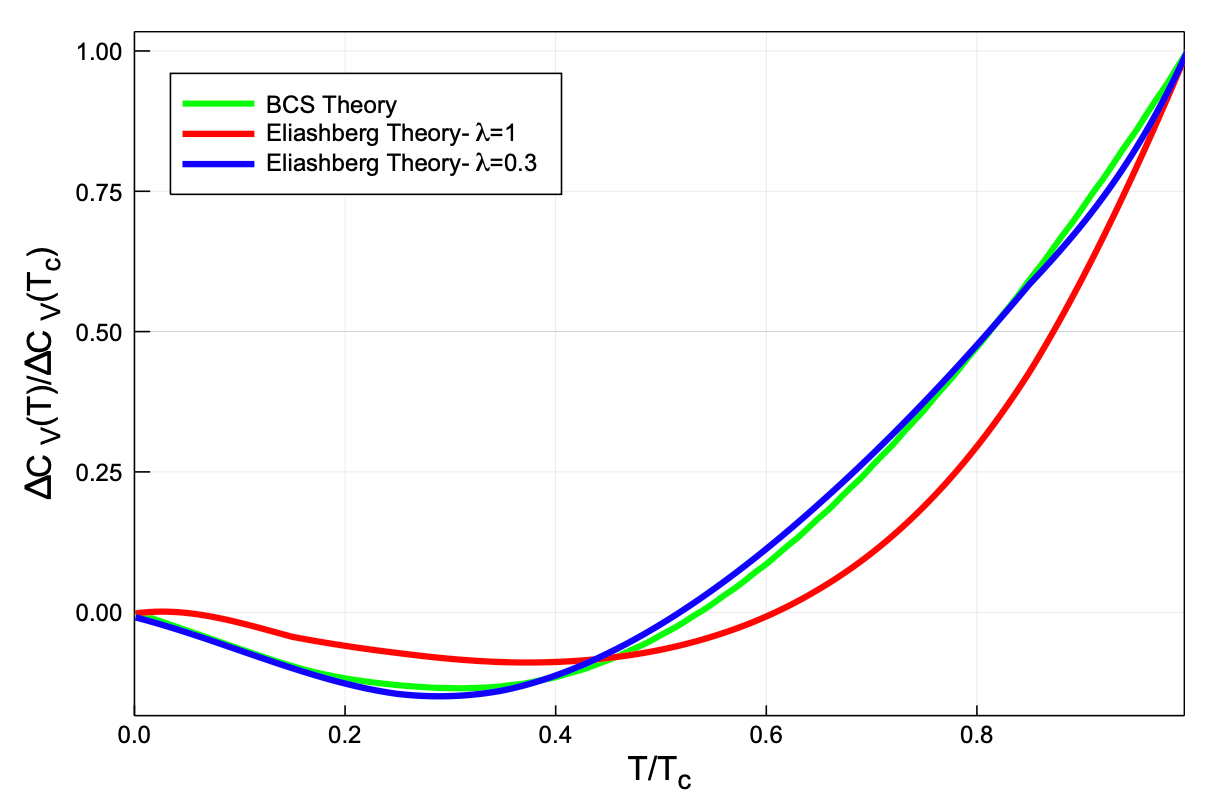}
\centering
\caption{A plot of $\Delta C_{V}(T)/\Delta C_{V}(T_{c})$ versus $T/T_{c}$ for the Eliashberg and BCS theories. The $\lambda=0.3$ weak-coupling Eliashberg plot (blue) and the BCS plot (green) are similar. However, the $\lambda=1$ Eliashberg plot (red) is different.}
\label{fig:SpecificHeat2}
\end{figure}

In Fig.~\ref{fig:SpecificHeat2} we plot $\Delta C_{V}(T)/\Delta C_{V,n}(T_{c})$ versus $T/T_{c}$ for the Eliashberg and BCS theories. For the case of $\lambda=0.3$, the weak-coupling Eliashberg theory curve is in good agreement with the BCS result. 
Again, we emphasize the fact that $\Delta C_{V}(T)$ for weak-coupling Eliashberg theory will differ by $1/\sqrt{e}$ from the BCS result: it is the dimensionless ratio of specific heats that is the same for the two theories. 
For small temperatures the $\lambda=0.3$ curve is below the $\lambda=1$ curve and both plots are negative, whereas for temperatures $T/T_{c}\gtrapprox0.6$, both curves change sign and the $\lambda=0.3$ curve lies above the $\lambda=1$ curve. The weights of these curves that resides above or below the temperature axis is constrained by the continuity of entropy~\cite{Rickayzen,AGD,FetterWalecka}. Indeed, entropy is related to specific heat by $C_{V}=\left.TdS/dT\right|_{V}$ and thus, if we integrate this equation with respect to temperature and use the third-law of thermodynamics, then we obtain $\Delta S(T)=\int_{0}^{T}dT^{\prime} \Delta C_{V}\left(T^{\prime}\right)/T^{\prime}$. The number of configurations of a system is a discrete quantity, and moreover it must be a continuous function of temperature; since entropy is proportional to the logarithm of the number of configurations, it follows that entropy is a continuous function of temperature and thus $\Delta S(T_{c})=0$. Hence, when the curves in Fig.~\ref{fig:SpecificHeat2} are divided by $T$ and the integral over temperature from $T/T_{c}=0$ to $T/T_{c}=1$ is performed, the result must vanish. This explains the sign change in the plots of $\Delta C_{V}$. 

\subsection{Critical magnetic field}

\begin{figure}[h]
\centering
\includegraphics[width=8.25cm]{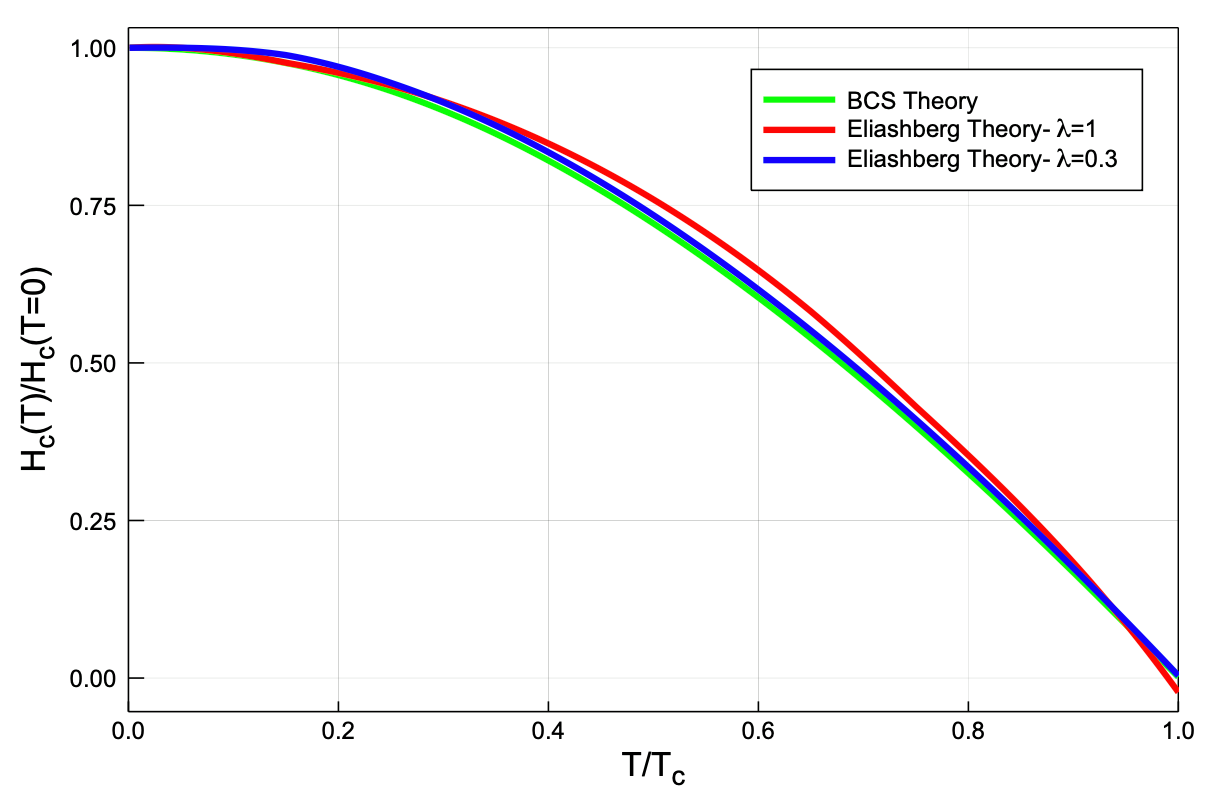}
\caption{A plot of the normalized critical magnetic field versus reduced temperature for Eliashberg theory with $\lambda=1$ (red), $\lambda=0.3$ (blue), and also for BCS theory (green).}
\label{fig:CriticalField1}
\end{figure}

The critical magnetic field is defined by~\cite{Bennemann,Rickayzen}
\begin{equation}
H_{c}=\sqrt{-8\pi \Delta F},
\end{equation}
where $\Delta F$ is the free-energy difference, which we determine via Eq.~\eqref{eq:FreeEnergy}. 
In BCS theory, the zero-temperature limit of the critical magnetic field is~\cite{Rickayzen,AGD}:
\begin{equation}
\label{eq:HCrit1}
\left.\frac{H_{c}(T)}{H_{c}(0)}\right|_{\text{BCS}} \rightarrow 1-e^{2\gamma/3}\left(\frac{T}{T_{c}}\right)^{2}, \quad T\rightarrow{0}.
\end{equation}
The numerical value of the coefficient in front of the $T^2$ term in the equation above is $e^{2\gamma/3}\approx1.06$, which is close to unity~\cite{AGD}. 
The zero-temperature critical magnetic field appearing above is defined by $H_{c}^{2}(0)/8\pi=N(0)\Delta_{0}^{2}/2$.

As $T\rightarrow T_{c}$, the critical magnetic field, as computed within BCS theory, is~\cite{AGD}:
\begin{equation}
\label{eq:HCrit2}
\left.\frac{H_{c}(T)}{H_{c}{(0)}}\right|_{\text{BCS}} \rightarrow e^{\gamma}\sqrt{\frac{8}{7\zeta(3)}}\left(1-\frac{T}{T_{c}}\right),\quad T\rightarrow T_{c}.
\end{equation}
The numerical value of the prefactor is $\approx1.74$~\cite{AGD}. In Fig.~\ref{fig:CriticalField1}, the normalized critical magnetic field $H_{c}(T)/H_{c}(0)$ versus  $T/T_{c}$ is shown for Eliashberg theory, for $\lambda=1$ and $\lambda=0.3$, and also for BCS theory, and there is good agreement between weak-coupling Eliashberg theory and BCS theory. Indeed, this figure shows that the normalized critical magnetic field ratio, computed using weak-coupling Eliashberg theory, is in good agreement with the asymptotic limits for BCS theory written in Eqs.~\eqref{eq:HCrit1}-\eqref{eq:HCrit2}.

\section{Conclusion}
\label{sec:Conclusion}
In this paper we have extended the understanding of weak-coupling Eliashberg theory by studying its thermodynamic properties. Combined with the previous weak-coupling analyses of the gap and renormalization parameters, the present complementary analysis of the free energy, specific heat, and critical magnetic field culminates in a thorough elucidation of the single-particle properties of weak-coupling Eliashberg theory. 
In particular, we applied the Bardeen-Stephen formula for the free-energy difference between the superconducting and normal states and used this to show that, 
in the weak-coupling limit, the free-energy difference of Eliashberg theory has a correction of $1/e$ in comparison to the BCS case, for all temperatures below the critical temperature.
Furthermore, we showed that, while there is a correction of $1/\sqrt{e}$ in the respective specific-heat difference in Eliashberg theory and BCS theory, when normalizing this quantity by $T_{c}$
we find agreement between these two theories. We also illustrated this agreement in the normalized weak-coupling properties of the critical magnetic field. 
Moreover, we showed that the discontinuity in the specific heat, as the temperature approaches the critical temperature, is the same for weak-coupling Eliashberg theory and BCS theory.
This provides credence to the notion that the ``universal" ratios in BCS theory are recovered in weak-coupling Eliashberg theory.

\acknowledgments
This work was supported in part by the Natural Sciences and Engineering Research Council of Canada (NSERC), and by an MIF from the
Province of Alberta. R.B. acknowledges support from the Department of Physics and the Theoretical Physics Institute at the University of Alberta.

\bibliographystyle{apsrev4-1}
\bibliography{SpecificHeat_Bibliography.bib}
\end{document}